\documentclass[iop]{emulateapj}

\newcommand{\Msun}{{\rm M_{\odot}}}

\newcommand{\Zsun}{{\rm Z_{\odot}}}
\newcommand{\Rsun}{{\rm R_{\odot}}}
\newcommand{\mpc}{\, {\rm Mpc}}

\newcommand{\Halpha}{{\rm H}{\alpha}}
\newcommand{\Ha}{{\rm H}{\alpha}}
\newcommand{\Rc}{{\rm R}_{\rm C}}
\newcommand{\NA}{{\rm NA659}}

\slugcomment{}

\shorttitle{Subaru $\rm  H_{\alpha}$ Observations of M83 XUV Disk}
\shortauthors{Jin Koda}

\begin{document}

\title{The Universal Initial Mass Function In The XUV Disk of M83}

\author{Jin Koda\altaffilmark{1},
Masafumi Yagi\altaffilmark{2},
Samuel Boissier\altaffilmark{3},
Armando Gil de Paz\altaffilmark{4},
Masatoshi Imanishi\altaffilmark{5},
Jennifer Donovan Meyer\altaffilmark{1},
Barry F. Madore\altaffilmark{6},
David A. Thilker\altaffilmark{7}}

\email{jin.koda@stonybrook.edu}

\altaffiltext{1}{Department of Physics and Astronomy, Stony Brook University, Stony Brook, NY 11794-3800}
\altaffiltext{2}{Optical and Infrared Astronomy Division, National Astronomical Observatory of Japan, 2-21-1 Osawa, Mitaka, Tokyo, 181-8588, Japan}
\altaffiltext{3}{Laboratoire d'Astrophysique de Marseille - LAM, Universit\'e Aix-Marseille \& CNRS, UMR7326, 38 rue F. Joliot-Curie, 13388 Marseille Cedex 13, France}
\altaffiltext{4}{Departamento de Astrof\'{\i}sica y CC. de la Atm\'{o}sfera, Universidad Complutense de Madrid, Avda. de la Complutense s/n, Madrid E-28040, Spain}
\altaffiltext{5}{Subaru Telescope, 650 North A'ohoku Place, Hilo, HI 96720}
\altaffiltext{6}{Carnegie Institution of Washington, 813 Santa Barbara Street, Pasadena, CA 91101}
\altaffiltext{7}{Center for Astrophysical Sciences, The Johns Hopkins University, 3400 N. Charles Street, Baltimore, MD 21218}

\begin{abstract}
We report deep Subaru $\Halpha$ observations of the extended ultraviolet (XUV) disk
of M83\footnote{Based in part on data collected at Subaru Telescope, which is operated
by the National Astronomical Observatory of Japan.}.
These new observations enable the first complete census of very young stellar clusters over the entire XUV disk.
Combining Subaru and GALEX data with a stellar population synthesis model,
we find that 
(1) the standard, but stochastically-sampled, initial mass function (IMF) is preferred over the truncated IMF,
because there are low mass stellar clusters ($10^{2-3}\Msun$) that host massive O-type stars; that
(2) the standard Salpeter IMF and a simple aging effect explain
the counts of FUV-bright and $\Ha$-bright clusters with masses $>10^3\Msun$; and that
(3) the $\Ha$ to FUV flux ratio over the XUV disk supports the standard IMF.
To reach conclusion (2), we assumed instantaneous cluster formation and
a constant cluster formation rate over the XUV disk.
The Subaru Prime Focus Camera (Suprime-Cam) covers a large area even
outside the XUV disk -- far beyond the detection limit of the HI gas.
This enables us to statistically separate the stellar clusters in the disk from background
contamination.
The new data, model, and previous spectroscopic studies provide overall consistent results
with respect to the internal dust extinction ($A_{\rm V}\sim 0.1$ mag) and low metallicity ($\sim 0.2\Zsun$)
using the dust extinction curve of the Small Magellanic Cloud.
The minimum cluster mass for avoiding the upper IMF incompleteness due to stochastic
sampling and the spectral energy distributions of O, B, and A stars are discussed in the Appendix.
\end{abstract}


\section{Introduction}

The initial mass function (IMF) is of central importance in modeling galaxy formation and evolution.
Its universality has been supported observationally \citep[e.g., ][]{kro02, bas10} and adopted in many applications \citep[e.g., ][]{ken98}.
Observations of resolved stellar populations and integrated colors of most galaxies are consistent with a universal IMF,
but a non-universal IMF has been inferred recently, possibly depending on environmental parameters \citep{bas10}.
The IMF could be steeper (or truncated), as compared to the standard one, in low density environments,
such as  dwarfs and low surface brightness (LSB) galaxies, thus producing
fewer high-mass stars \citep{hov08, meu09, lee09}.
On the other hand, a top heavy IMF is suggested in dense starburst galaxies \citep{rie93, par07}.

\citet{pfl08} explained the apparent non-universality as a purely statistical effect; 
the intrinsic IMF shape is invariant, but the maximum mass of the stars in a stellar cluster
is limited by the mass of the parental cluster, the mass of which is in turn controlled by star formation activity.
Thereby, the IMF integrated over a galaxy -- integrated galactic IMF (IGIMF), may apparently differ
in  low density environments.
\citet{kru08} suggested another model: a low gas density could result in fragmentation into 
gas clumps too small to allow high-mass star formation.
The non-universality of the IMF potentially alters our understanding of star formation and galaxy evolution.
Its confirmation is urgent. 

The Galaxy Evolution Explorer (GALEX) provides a means to investigate the high-mass end of the  IMF.
The combination of GALEX and $\Halpha$ data can separate the populations of O and B stars;
the two GALEX bands, FUV (1539$\rm \AA$) and NUV (2316$\rm \AA$), are sensitive to
both stellar clusters with O stars and with B stars,
while $\Halpha$ data are responsive predominantly to clusters with O stars.
\citet{meu09} analyzed the $\Halpha$ to FUV flux ratio over galaxies,
found a deficit of ionizing O stars, and concluded that the IMF is steeper (or truncated) in dwarfs and LSBs.
This analysis, however, is sensitive to the procedure of dust extinction correction
\citep[see alternative analysis; ][]{bos09}.
Another approach, the comparison of $\Halpha$ flux with cluster mass, points to a universal IMF \citep{cal10}.

GALEX finds UV emission far beyond the optical edge of galaxies,
i.e., their extended ultraviolet disks \citep[XUV disks; ][]{thi05, gil05, thi07}.
Star formation in such low density outskirts is a prime target to study the IMF variation.
In fact, the radial UV profile extends beyond the $\Ha$ truncation radius \citep{boi07},
indicating that O stars are extremely rare while B stars are abundant.
The absence may be an indication of a non-universal IMF,
or could simply be the effect of mixed
stellar ages \citep{thi05} -- as O stars are the primary sources of HII regions and
have much shorter lives ($\sim$ 6 Myr) than B stars ($\sim$ 100 Myr).
Indeed, \citet{god10} averaged $\Halpha$ and FUV fluxes over XUV disks and found
the ratio consistent with the prediction from the standard IMF, although the 
sensitivity and area coverage of their $\Halpha$ data were limited. 

We present $\Halpha$ observations of M83 obtained with the Subaru Prime-focus Camera
\citep[Suprime-Cam; ][]{miy02} on the Subaru telescope \citep{iye04}.
With its high sensitivity and wide field-of-view ($34\arcmin \times 27\arcmin$),
we can cover the entire XUV disk and surrounding area with two pointings
(Figure \ref{fig:coverage}) and can detect HII regions around individual massive stars.
M83 is one of the nearest galaxies with a prominent XUV disk \citep[$d=4.5\mpc$ or $m-M=28.27$ mag; ][]{thi03}
and is best suited for the detection of individual HII regions.
This study provides the first complete census of the UV-bright cluster population
with and without HII regions.
We count the number of blue clusters in the UV and bright clusters in $\Ha$ and translate
the number ratio into the ratio of durations when a cluster is UV-blue and $\Ha$-bright.
We compare the ratio with the prediction from a stellar population synthesis model using
the standard IMF.

\section{Data}
\subsection{Subaru}

\begin{figure}
\epsscale{1.0}
\plotone{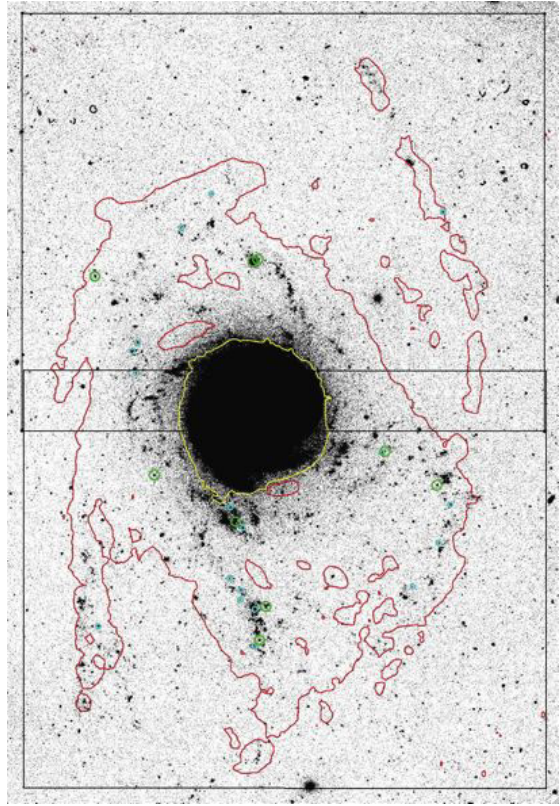}
\caption{
Subaru coverage and definition of regions in M83. The background is a GALEX FUV image downloaded from {\sc Galexview}. Solid overlapping rectangles are two pointings with the Subaru Prime Focus Camera (Suprime-Cam; the field-of-view is $34\arcmin \times 27\arcmin$, or $44.5 \times 35.3\,\rm kpc^2$,  in one pointing). The inner (yellow) contour is roughly at the traditional edge of the optical disk. The outer (red) counter is at an HI surface density of $1.5\times10^{20} \rm \, cm^{-2}$, using the HI data from \citet{mil09}. We call FUV-bright objects inside the HI contour IN objects (excluding those in the inner disk - yellow contour), and those outside OUT objects. The H$\alpha$-bright clusters ($\NA-\Rc<-1$ mag; \S \ref{sec:sample}) are also plotted: green for clusters with $>10^3\Msun$ and cyan for the ones with $<10^3\Msun$.
\label{fig:coverage}}
\end{figure}

Observations were made in the $R_C$ and $\NA$ ($\Halpha$) bands \citep{oka02, yos02, hay03} on September 7, 2010.
Two Suprime-Cam fields cover the entire XUV disk of M83, i.e., the whole area with an HI surface density above
$1.5\times 10^{20} \,\rm cm^{-2}$ \citep[roughly the detection limit in ][]{mil09} and a large surrounding area.
Figure \ref{fig:coverage} shows the area coverage and the HI contour (red) on a GALEX FUV image.
The two fields are marked with black overlapping solid boxes.
The exposure time for each field is 5$\times$12 minutes and 5$\times$4 minutes in the $\NA$ and $\Rc$
bands, respectively.
The yellow contour indicates the edge of the optical disk (roughly, the 25 mag/arcsec$^2$
isophote). The typical seeing was $\sim 1\arcsec$.

Data were reduced in a standard way using the Suprime-Cam reduction software \citep[SDFRED; ][]{yag02, ouc04}.
We performed overscan subtraction, flat-fielding, distortion correction, background subtraction, and mosaicking.
The flux calibration was performed against standard field stars \citep{oke90, lan92}.
$R_C$ and $\NA$ bands are practically at the same wavelength \citep{hay03},
and we do not expect any
spatial variation in the $\Rc/\NA$ ratios of typical foreground field stars.
However, there remain residual errors at large spatial scales in the ratios (a few perent)
which were likely introduced in flat fielding.
We made a correction by making a $\NA/\Rc$ ratio map using the average ratio of field stars
in each $200\arcsec \times 200\arcsec$ grid and by normalizing the $\NA$ image.
The $1\sigma$ photometric limits with the $\sim 5\arcsec$ aperture (i.e., GALEX resolution)
are 24.90 and 25.56 AB mag in the $\NA$ ($\Halpha$) and $R_C$, respectively. 

\subsection{GALEX}

We use archival GALEX far-ultraviolet (FUV) and near-ultraviolet (NUV) images \citep{big10}.
The data are obtained from the GALEX data archive through {\sc Galexview} in the multimission
archive at STScI (MAST). The GALEX observations and data reduction were discussed in \citet{big10}.
The FWHM sizes of the point spread functions (PSF) are measured with the IDL software ATV \citep{bar01} and are
$4.7\arcsec$ for FUV and $5.2\arcsec$ for NUV.
The $1\sigma$ photometric limits with the $\sim 5\arcsec$ aperture are 26.34 and 26.20 AB mag
in the FUV and NUV-bands, respectively. 

\section{Methods}


\subsection{Stellar Population Synthesis Model}\label{sec:model}

Our analysis is guided by the  stellar population synthesis model {\sc Starburst99} \citep{lei99}.
The Padova stellar evolution tracks with AGB stars are adopted.
Most UV-bright objects in the XUV disk of M83 are likely stellar clusters \citep{thi05, gil07}. Therefore, we adopt
an instantaneous burst approximation to model their photometric evolution.

\begin{figure}
\epsscale{1.0}
\plotone{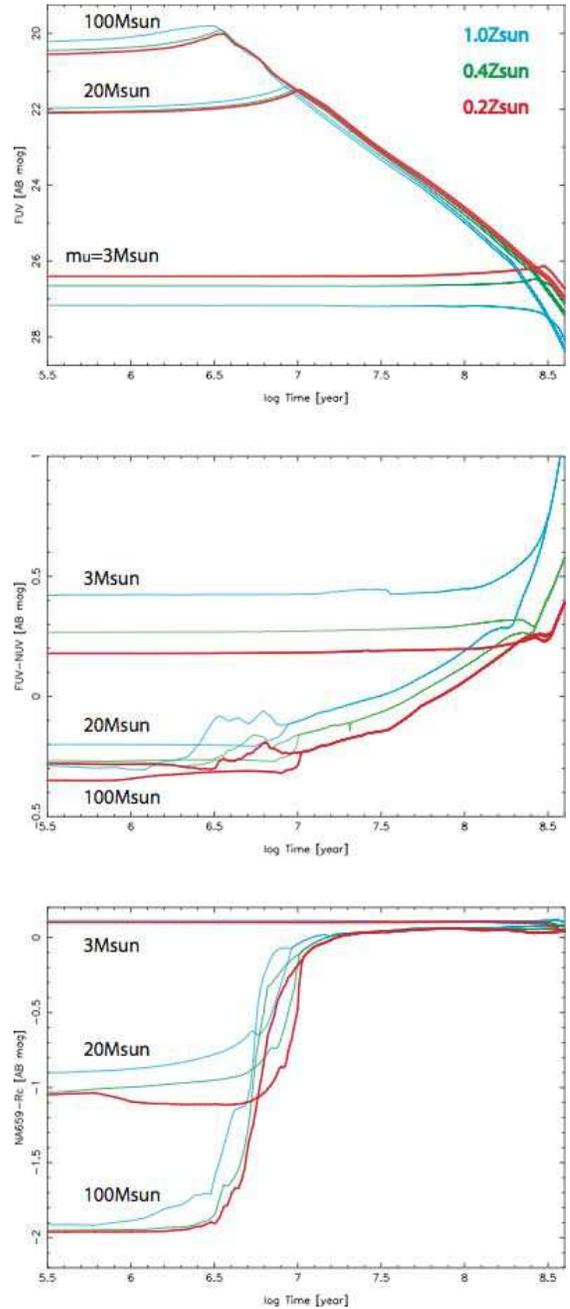}
\caption{Photometric evolution of stellar clusters from {\it Starburst99}. A single-burst model with the Salpeter IMF is adopted. A low metallicity (0.2 $\Zsun$) is adopted (red) for the analysis, but models of higher metalicities (0.4$\Zsun$ and 1.0$\Zsun$) are also plotted (green and cyan, respectively). The lower mass boundary for the IMF is $0.1\Msun$, and the upper mass boundary is $100\Msun$ (including all types of stars), $20\Msun$ (no O stars), and $3\Msun$ (no OB stars). The adopted distance of M83 is $d=4.5\mpc$.
{\it Top:} FUV magnitude for a cluster mass of $1000\Msun$.
{\it Middle:} FUV-NUV color.
{\it Bottom:} $\NA$-$R_C$ color.
\label{fig:model}}
\end{figure}

We will show that the standard Salpeter IMF \citep{sal55} explains the observations well.
Figure \ref{fig:model} demonstrates the differences between 
the standard and truncated IMFs.
The cluster mass is set to $1000\Msun$ for the plot, which is roughly the minimum mass of a cluster
to fully populate the IMF up to an O star (see Appendix \ref{sec:mcl}).
Note that a factor of 10 increase in cluster mass shifts all the magnitude curves in Figure \ref{fig:model} (top)
systematically upward by 2.5 magnitudes, but it does not change the color curves.
For these models, we assume the Salpeter IMF with a lower cut-off mass of
$m_l=0.1\Msun$. The upper cut-off mass is set to $m_u=$ 100, 20, and 3 $\Msun$,
which corresponds to the masses of the most massive O, B, and A-type stars, respectively \citep{cox99}.
In other words, the last two models represent the truncated IMF (or, almost equivalently, the steeper IMF)
without O or OB stars.
We adopt 20\% of the solar metallicity ($0.2\Zsun$) from the spectroscopic measurements \citep{gil07, bre09}.
For a reference, the spectra of individual O, B and A stars are discussed in Appendix \ref{sec:starspec}.

There are quantitative differences between clusters with and without OB stars, i.e., between
the standard and truncated (steeper) IMFs.
Clusters without OB stars are fainter by 6 mag in FUV for the first 10 Myr and
never become as blue as FUV-NUV$<$0.2$-$0.3 mag. Therefore, blue clusters ($<$0.2$-$0.3 mag)
must have O and/or B stars.
The FUV-NUV color is insensitive to the difference between clusters with only B stars (up to
$m_u=20\Msun$) and with both O and B stars ($100\Msun$).
This is primarily because the GALEX bands trace only the Rayleigh-Jeans side of stellar thermal
radiation for both O and early-type B stars (Appendix \ref{sec:starspec}). The FUV-NUV color is therefore a useful indicator
of the presence of O and/or B stars in a cluster.

The $\NA-\Rc$ color differentiates clusters with O stars from those without them, since O stars emit
a significantly greater number of ionizing photons than B stars \citep{ste03}.
This is demonstrated in Figure  \ref{fig:model} (bottom).
Clusters with $\NA-\Rc <$  -1 mag must have O stars, and those with $\NA-\Rc <$  -0.3 mag should have
O and/or B stars.
For this plot, we calculated the $\Ha$ line luminosity $L_{\Ha}$ using the Lyman photon flux $Q_{\rm Lyc}$
from {\sc Starburst99} and the relation $L_{\Ha} {\rm [erg \, s^{-1}]}= 1.37 \times 10^{-12}Q_{\rm Lyc} {\rm [s^{-1}]}$
from case B recombination with an electron temperature of $10^4$K \citep{gav02}.
Case B recombination is perhaps a reasonable assumption even in the low-density environment,
since the optical depths in the Lyman resonance lines are very likely quite large \citep{ost06}
in any star-forming conditions.
The $\NA$ magnitude is based on the stellar continuum luminosity plus $L_{\Ha}$.
We ignore the effect of [NII] emission since the effect is very small \citep[][ see \S \ref{sec:fuvsel}]{gil07, ken08, god10}.

\subsection{Cluster Counts}
  
Counting blue clusters in FUV-NUV and $\NA-\Rc$ constrains the IMF at its high mass end.
Clusters become bright in $\Halpha$ (``blue" in the color $\NA-\Rc$) predominantly due to O stars,
while FUV-NUV is due to both O and B stars.
Since O stars have much shorter lives than B stars, $\NA-\Rc$ changes faster than FUV-NUV (Figure \ref{fig:model}).
Under the assumptions of  instantaneous cluster formation and a constant cluster formation rate,
the numbers of blue clusters in FUV-NUV and $\NA-\Rc$ ($N_{\rm UV}$ and $N_{\Halpha}$) should
be proportional to the durations that clusters are blue ($t_{\rm UV}$ and $t_{\Halpha}$, respectively).
Thus, we expect $N_{\Halpha}/N_{\rm UV} = t_{\Halpha}/t_{\rm UV}$ if the difference in the populations
is due to a simple aging effect starting from the initial populations of the standard IMF.
The assumption of a constant cluster formation rate should be approximately accurate
when integrated over the large XUV disk and over a short timescale, i.e., the young ages
of clusters which can be detected with our color selection criteria (below).
In fact, the clusters that we count in \S \ref{sec:uniimf} are distributed over the UV disk (Figure \ref{fig:coverage}).

For the standard IMF ($m_u=100\Msun$ and $0.2\Zsun$),
a cluster is blue (e.g., FUV-NUV$<0.0$ mag) for $t_{\rm UV} = 70.8$ Myr
and $\Ha$-bright (e.g., $\NA-\Rc<-1.0$ mag) for $5.8$ Myr.
Therefore, the number ratio should be $N_{\Halpha}/N_{\rm UV} = 0.08$
if each population is the consequence of the cluster aging effect.
We will compare the observations with this theoretical prediction.

We identify and count stellar clusters at the GALEX resolution
($\sim 5\arcsec$; $\sim 109$ pc at the distance of M83).
It is possible that  clusters may be blended at this resolution; however,
even in that case, our analysis is valid if clusters formed in such proximity
are typically coeval.

\section{Sample Selection}\label{sec:sample}

\subsection{FUV-Selected Objects}\label{sec:fuvsel}
UV-bright objects are identified based on the FUV image using the {\sc SExtractor} package \citep{ber96}.
Saturated stars and image edges are masked before the identification.
Objects within the traditional optical disk (yellow contour in Figure \ref{fig:coverage}) are excluded.
We confirmed the very low probability of false detection using the negative image technique.
{\sc SExtractor} is run against the sign-flipped FUV image and detected
only one false target at 25.41 mag in FUV. Thus, false detections have a negligible
impact on our number counts.

The aperture photometry is carried out in the four bands with the dual image mode of {\sc SExtractor}
using the FUV image as the reference.
The $1\sigma$ photometric limits with the aperture of the GALEX resolution
($\sim 5\arcsec$) are 24.90, 25.56, 26.34, and 26.20 mag in the $\NA$ ($\Halpha$),
$R_C$, FUV, and NUV-bands, respectively. The sensitivity in $\NA$
(FWHM=120$\rm \AA$) corresponds to a flux of $7.98\times 10^{34} \rm \, erg \, s^{-1}$
at a distance $d=4.5\mpc$, which is sufficient to detect an HII region around
a single B star (e.g., $3.44\times 10^{35} \rm \, erg \, s^{-1}$ in case of B0; assuming case B recombination;
see Appendix \ref{sec:starspec}).
We note that the $\NA$ band includes stellar continuum emission as well as $\Halpha$ emission.

The contamination from [NII]$\lambda6548$, $6583$ in the $\NA$ photometry is neglected.
The line ratio [NII]/$\Halpha$ is small ($\lesssim 0.1$) in the outskirts of galaxies
\citep{gil07, ken08, god10}.  Even in the extreme case that the stellar continuum has zero contribution
to the $\NA$ flux, the error is only 0.1 mag (and likely less in more realistic conditions).

Figure \ref{fig:coverage} shows that the Subaru data cover both the inside and outside
of the HI contour (roughly the HI detection limit). 
Recently-formed FUV-bright objects  are more likely to reside inside the HI contour
as star formation occurs in gas.
For the background subtraction in \S \ref{sec:background}, the objects inside the HI contour (hereafter,
IN objects) and outside (OUT objects) are separately cataloged.
Note that we intensively used the IMCAT software package (Nick Kaiser, private communication)
\footnote{See http://www.ifa.hawaii.edu/$\sim$kaiser/imcat/.} for the following data analysis.

In \S \ref{sec:uniimf}, we will discuss a selection of clusters based on their $\NA-\Rc$ color;
but in advance, Figure \ref{fig:coverage} shows the locations of clusters with $\NA-\Rc<-1$ mag (i.e., bright in $\Ha$).
They are distributed over the entire XUV disk, and all are within the HI contour (see \S \ref{sec:background}).

\subsection{Extinction Correction}\label{sec:extcor}
The extinction curves of \citet{pei92} are adopted to correct for Galactic and internal extinctions.
The extinctions in FUV, NUV, $R_{\rm C}$, and $\NA$ are
($A_{\rm FUV}$, $A_{\rm NUV}$, $A_{R_{\rm C}}$, $A_{\NA}$) = (2.58, 2.83, 0.81, 0.81)$A_{\rm V}$
for the Galaxy, (3.37, 2.61, 0.80, 0.80)$A_{\rm V}$ for the Large Magellanic 
Cloud (LMC), and (4.18, 2.59, 0.79, 0.79)$A_{\rm V}$ for the Small Magellanic Cloud (SMC),
where $A_{\rm V}$ is the extinction in $V$-magnitude.
We note that the FUV-NUV color becomes bluer under the Galactic extinction,
but redder under the LMC and SMC extinction.

The Galactic extinction correction is made for all objects using $A_{\rm V}=0.218$ mag \citep{sch98}.
We do not apply an internal extinction correction until \S \ref{sec:discussion},
but it will be corrected using $A_{\rm V}=0.1$ mag and the SMC extinction curve
for the following reason.
The measured metallicity of most outer objects is low \citep[$\sim 0.2\Zsun$; ][]{gil07},
and therefore, the extinction curve of the SMC is perhaps the most appropriate.
Using $E(B-V)$ values from \citet{gil07}, we estimate the amount of internal extinction to be
around $A_{\rm V} \sim 0.1$ mag with outliers (e.g., 0.0 - 1.0 mag).
We therefore adopt $A_{\rm V}\sim 0.1$ mag for the statistical correction,
and this degree of extinction will be supported by a comparison with
a model color in \S \ref{sec:intext2}.

\subsection{Completeness}\label{sec:comp}
\begin{figure}
\epsscale{1.0}
\plotone{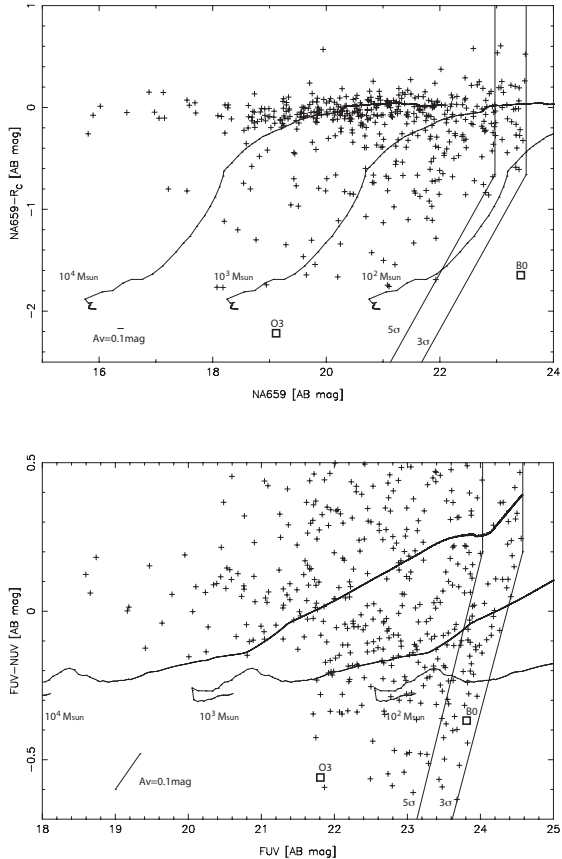}
\caption{Color-magnitude diagrams of all the objects inside the HI contour (IN objects defined in \S \ref{sec:fuvsel}).
{\it Top:} $\NA$-$\Rc$ vs $\NA$. $5\sigma$ and $3\sigma$ detection limits are illustrated with solid straight lines.
The model evolutionary sequences with the standard Salpeter IMF are displayed with solid curves for clusters with $10^2$, $10^3$, $10^4\Msun$.
{\it Bottom:} The same as the top panel, but for FUV-NUV vs FUV.
Single O3 and B0 star models are also plotted in both panels.
\label{fig:colormag}}
\end{figure}

The mass detection limit is determined in color magnitude space with the help of models.
Figure \ref{fig:colormag} shows color-magnitude diagrams for IN objects.
The overlaid solid lines are the 3$\sigma$ and 5$\sigma$ magnitude detection limits and model
evolutionary sequences for clusters with masses of $10^2$, $10^3$, and $10^4\Msun$.
The models are for clusters with the standard Salpeter IMF with ($m_l$, $m_u$) = (0.1, 100$\Msun$)
and metallicity $0.2\Zsun$.
Models with lower $m_u$ do not reproduce clusters as blue as observed (Figure \ref{fig:model};
and see discussion below).
Clearly, the mass detection limit changes as a function of color in Figure \ref{fig:colormag};
it decreases as clusters become redder.
We note that the $\Rc$ and $\NA$ bands are practically at the same wavelength,
and therefore, $\NA-\Rc$ is zero without $\Ha$ emission.
A negative $\NA-\Rc$ color indicates the presence of associated $\Ha$ emission.

The top panel shows that the Subaru observations detect clusters down to $\sim 10^2\Msun$.
Clusters become redder and fainter as they age, and those around the detection
boundary ($10^2\Msun$) become too faint when they become redder than $\NA-\Rc\sim -0.6$ mag.
In other words,  the sample is complete down to $10^2\Msun$ in the range $\NA-\Rc < -0.6$ mag.
It is complete down to $10^3\Msun$ in $\NA-\Rc<0.0$ mag.
The extinction is negligible, as shown by the extinction vector (for $A_{\rm V}=0.1$ mag) in the plot.

The bottom panel shows the corresponding color-magnitude diagram in the UV.
Again, the mass detection limit is a function of the color range (and thus, the cluster age).
The detection limit is $\sim 10^3\Msun$ for the color  FUV-NUV$<0.0$ mag
and is even deeper ($\sim 10^2\Msun$) for FUV-NUV$<-0.2$ mag.
The extinction is not negligible in FUV and NUV and will be discussed later.

Note that the very blue objects in the bottom panel (FUV-NUV$\lesssim -0.3$ mag
and FUV $\sim 22$ - 24 mag) are bluer than any model cluster
with the standard IMF can reproduce, and are located approximately in the region
where single O stars appear (the models of O3 and B0 stars are also plotted).
These objects are not included in our number count, since they appear below
the mass selection criterion in \S \ref{sec:uniimf}.
We will discuss these objects in \S \ref{sec:trusto}.

\subsection{Background Subtraction}\label{sec:background}

The contamination of background (and foreground) objects has been an obstacle in identifying
the objects within nearby galaxies \citep{wer10, god10}.
The large field-of-view of Suprime-Cam is a clear advantage as it enables us to remove such contamination statistically.
The objects outside the HI gas contour (OUT objects; see \S \ref{sec:fuvsel}) are most likely in the background,
whereas IN objects include stellar clusters within M83 as well as background objects.

\begin{figure}
\epsscale{1.0}
\plotone{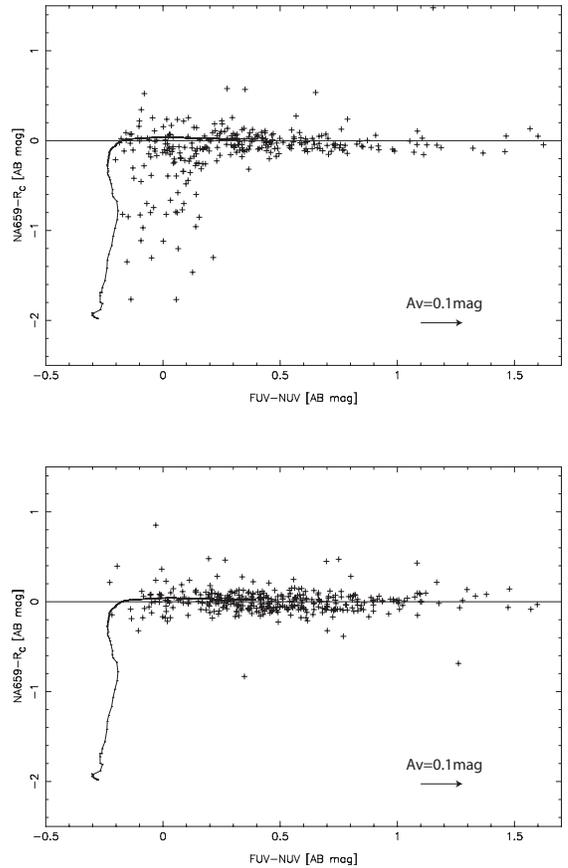}
\caption{$\NA$-$\Rc$ vs FUV-NUV color-color plot of the objects above $5\sigma$-detection and above $10^3\Msun$. Lines are model predictions with the Salpeter IMF with $m_u=100\Msun$ for 0.2 $\Zsun$. A reddening vector of $A_{\rm V}=0.1$ mag from the SMC extinction curve is also plotted. We assume that stellar clusters (traced in $\Rc$) and the surrounding ionized gas (in $\NA$)
suffer from extinction by the same amount.
{\it Top:} FUV-bright objects within the HI contour (IN objects). {\it Bottom:} the FUV-bright objects outside (OUT objects).
\label{fig:colorcolor}}
\end{figure}

\begin{figure}
\epsscale{1.1}
\plotone{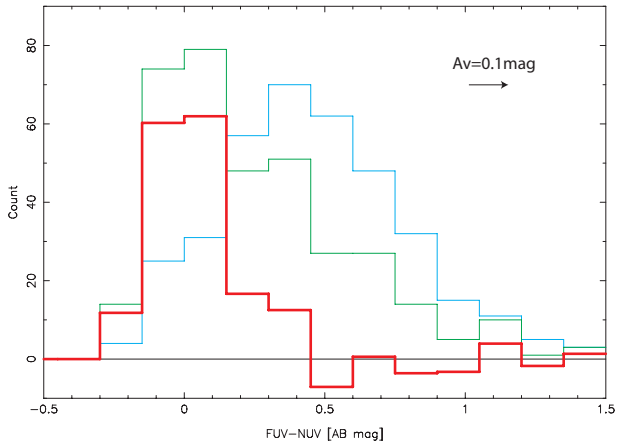}
\caption{Histograms of FUV-NUV color of the objects above $5\sigma$-detection and above $10^3\Msun$.
Colors indicate the objects within the HI gas (green), the ones outside of the HI gas (background objects; blue),
and the ones associated with M83 (red). The associated ones are calculated by subtracting OFF from ON
with scaling for the area coverage difference.
\label{fig:FUVhist}}
\end{figure}

The IN and OUT objects show different colors statistically.
Figure \ref{fig:colorcolor} shows the plots of $\NA$-$\Rc$ vs FUV-NUV for the IN and OUT objects
above the $5\sigma$ magnitude detection limit and above the $10^3\Msun$ threshold line
in Figure \ref{fig:colormag} (bottom).
Figure \ref{fig:FUVhist} is a histogram of FUV-NUV color for the IN (green) and OUT objects (blue).
Clearly, the IN objects are bluer in FUV-NUV than the OUT objects and preferentially
have FUV-NUV$<$0.2$-$0.3 mag, while the OUT objects are more likely to have FUV-NUV$>$0.2$-$0.3 mag.
Figure \ref{fig:colorcolor} shows that virtually all objects with $\NA-\Rc$ $<$ -0.3 mag
(OB stars) are IN objects; almost no young stellar clusters exist outside of
the HI gas. This clear distinction ensures that the objects outside of the HI gas
are not the young clusters in M83 and represent the background population.
Note that IN objects are young stellar clusters, not very old clusters
with many evolved FUV-bright stars, such as extreme horizontal branch stars, blue stragglers
or planetary nebulae.
The old clusters do not become as blue in FUV-NUV as our IN objects \citep{don09}.

The background contamination in the IN objects can be removed statistically;
we subtract the OUT objects from the IN.
In Figure  \ref{fig:FUVhist} the number of the OUT objects are scaled by the IN/OUT area ratio (55\%)
and subtracted from that of the IN objects.
The derived histogram (red) represents the population of young clusters with mass $>10^3\Msun$
in the XUV disk.

Virtually all clusters after the subtraction show blue color in FUV-NUV ($<$ 0.2$-$0.3 mag; Figure \ref{fig:FUVhist}),
suggesting that all of these young clusters have O or, at least, B stars (see Figure \ref{fig:model}).
Thus, massive stars are populated even in the diffuse gas environment.
It is also remarkable that almost no red objects (FUV-NUV $>$0.2$-$0.3 mag) exist.
This confirms that OUT objects are distributed everywhere across our field-of-view and
represent the background objects; thus, our background subtraction is fairly successful.

There is a possibility that old stellar clusters (i.e.,  FUV-NUV $>$0.2$-$0.3 mag before extinction correction)
exist both in the IN and OUT regions and are removed from the IN population.
However, it should not affect our cluster count analysis as we discuss only the young clusters.
Such a population, if it exists, must be fairly uniformly distributed both in the IN and OUT regions,
since the count after the subtraction is nearly zero in FUV-NUV $>$0.2$-$0.3 mag.
We also note that the apparent magnitudes of low mass clusters ($\lesssim 10^3\Msun$)
fall below our detection limit after $\sim100$ Myr (Figure \ref{fig:model} {\it top}).
Again, this is not a problem with the selection criteria that are adopted in our analysis (\S \ref{sec:uniimf}).

\section{Discussion}\label{sec:discussion}

\subsection{Truncated vs Stochastic IMFs}\label{sec:trusto}

The presence of ionizing stars (O stars) is evident even in small clusters ($<10^3\Msun$)
in Figure \ref{fig:colormag}, suggesting that the IMF is not truncated.
Massive stars do not form under the truncated IMF at all.
On the other hand, they should appear occasionally even in small clusters under the stochastic IMF,
though they tend not to appear in a small sample due to their low probability of formation in small clusters.
Such a stochastic effect becomes particularly apparent when the cluster mass is small ($<1000\Msun$).
Our Subaru observations detect small clusters (down to $\sim 10^2\Msun$)
and confirm that even those small clusters show $\Ha$ excess (i.e., $\NA-\Rc<0.0$ mag);
and some even have $\NA-\Rc<-1.0$ mag, a color which only O stars can produce.
Thus, O stars are present in the $10^{2-3}\Msun$ clusters.
The IMF is not truncated even in the low-density and low-metallicity environment.

More precisely, it is necessary to account for the stochastic effects in discussions of
very low mass clusters ($<10^{3} \Msun$) --  and even then, our result  still
supports the stochastic IMF over the truncated IMF, as we discuss below.
We first need to consider the likelihood that a cluster with a given small mass has a massive star.
In Appendix \ref{sec:mcl} we estimate that a $10^3\Msun$ cluster is likely to have as massive star as $40\Msun$;
the cluster mass must be greater than $\sim400\Msun$ to have a high probability of hosting at least one O star ($>20\Msun$);
a $10^2\Msun$ cluster likely hosts up to a $\sim 10\Msun$ star.

The $\NA-\Rc$ color is sensitive to the mass of the most massive star in a cluster when the cluster is young,
whereas FUV-NUV does not distinguish O stars from B stars well (\S \ref{sec:model}; Figure \ref{fig:model};
see also Appendix \ref{sec:starspec}).
If a cluster has no massive star, it should appear redder ($\NA-\Rc>$ -1 mag; Figure \ref{fig:model})
than the observed clusters  in the range $10^{2-3} \Msun$; yet some clusters appear in the range
$\NA-\Rc =$ -1$\sim$ -2 mag.
Massive O stars in small clusters should be populated stochastically.
In addition, a stochastically-populated single O star can outshine a cluster with a small mass
($10^{2-3} \Msun$; Figure \ref{fig:model}).
This may result in an overestimation of the cluster mass when we apply the standard IMF
with $m_u=100\Msun$ to the cluster (as done in Figure \ref{fig:colormag}).
Therefore, the masses of small clusters
that appear in the range $10^{2-3} \Msun$ in Figure \ref{fig:colormag} could be actually
even less. This prefers the stochastic IMF, as even the less massive clusters
have O stars.

The discussion so far is about clusters of stars. Another possible evidence for the presence
of O stars is the population of very blue objects (FUV-NUV$\lesssim-0.3$ mag)
in Figure \ref{fig:colormag} ({\it bottom}).
They have the apparent magnitude and color similar to those of a single O star.
They are possibly isolated, single O stars or small clusters with an O star(s)
but without many low-mass stars. Higher resolution images are necessary
to confirm them; their presence, if confirmed, may be stronger evidence
for O stars that are formed in the low-density environments.

\subsection{Internal Extinction Correction}\label{sec:intext2}

The internal extinction is assumed to be $A_{\rm V}=0.1$ mag from  spectroscopic
measurements \citep[\S \ref{sec:extcor}; ][]{gil07}. This degree of  extinction on average
is also supported by a comparison with the model curve in Figure \ref{fig:colorcolor} (top).
The extinction vector, corresponding to $A_{\rm V}=0.1$, is also plotted.
An extinction correction by this amount creates agreement between the UV data
and the model curve, providing additional support for the adopted $A_{\rm V}$.

We assume that a stellar cluster (traced in $\Rc$) and the surrounding ionized gas
(in $\NA$) suffer from extinction by the same amount.
The two components may have different spatial distributions and
a slightly different attenuation \citep[e.g., $E(B-V)_{\rm continuum}\sim0.44 E(B-V)_{\rm gas}$
in case of more active star-forming environments; ][]{cal01}.
However, the difference is negligible for our analysis since the extinction is small ($A_{\rm V}=0.1$).

\subsection{Universal IMF}\label{sec:uniimf}
Detected clusters with a color of $\NA-\Rc<-1$ mag are evidence that some clusters host O stars.
Here we test whether their abundance, relative to the ones without O stars,
can be explained by a simple aging effect with the standard IMF.
Under the assumptions of instantaneous formation of star clusters and
a constant cluster formation rate
over the XUV disk, the numbers and durations of blue clusters should
show the relation $N_{\Halpha}/N_{\rm UV} = t_{\Halpha}/t_{\rm UV}$,
if the relative cluster populations are due to the aging effect.

Consistent sample selection between GALEX and Subaru data is crucial in this analysis.
Each of the four photometric bands has its own detection limit (Figure \ref{fig:colormag}),
and a bias could be introduced if only magnitude detection limits are taken into account.
We therefore use the cluster mass and color as thresholds for cluster counts.
As discussed in \S \ref{sec:comp}, both GALEX and Subaru data are complete
at $\gtrsim 5\sigma$, in colors $\NA-\Rc<-0.3$ mag and FUV-NUV$<0.0$ mag
down to a cluster mass $10^3\Msun$.
This is true even after the internal extinction correction (\S \ref{sec:intext2})
if a slightly lower detection limit ($3.4\sigma$) is adopted.
The extinction correction is made for all four bands, which is equivalent to
shifting the data points along the extinction vectors in Figure \ref{fig:colormag}.
We select the mass limited sample after the extinction correction
using the technique discussed in \S \ref{sec:sample}.

Almost all of the clusters are in the range FUV-NUV $<$ 0 mag when the extinction
is taken into account (Figure \ref{fig:FUVhist}); thus, we adopt this color range for our analysis.
This is the range where  B stars definitely, and O stars possibly, exist in the clusters;
we count how many of them have O stars below.
Clusters are blue, FUV-NUV$<0$ mag, for only a short duration.
With the standard IMF of $(m_l, m_u)=(0.1, 100\Msun)$ and $0.2 \Zsun$, 
this is $t_{\rm UV} = 71$ Myr (Figure  \ref{fig:model}).
The short $t_{\rm UV}$ is advantageous, since the assumption of a constant cluster formation rate
is likely valid for a short period over the large area (i.e., the entire XUV disk)
-- no local events likely bias the statistics.
On the other hand, clusters with O stars  show $\NA-\Rc<-1$ mag for only 5.8 Myr.
The model therefore predicts $t_{\rm \Ha}/t_{\rm UV} \sim 0.08$.

\begin{figure}
\epsscale{1.1}
\plotone{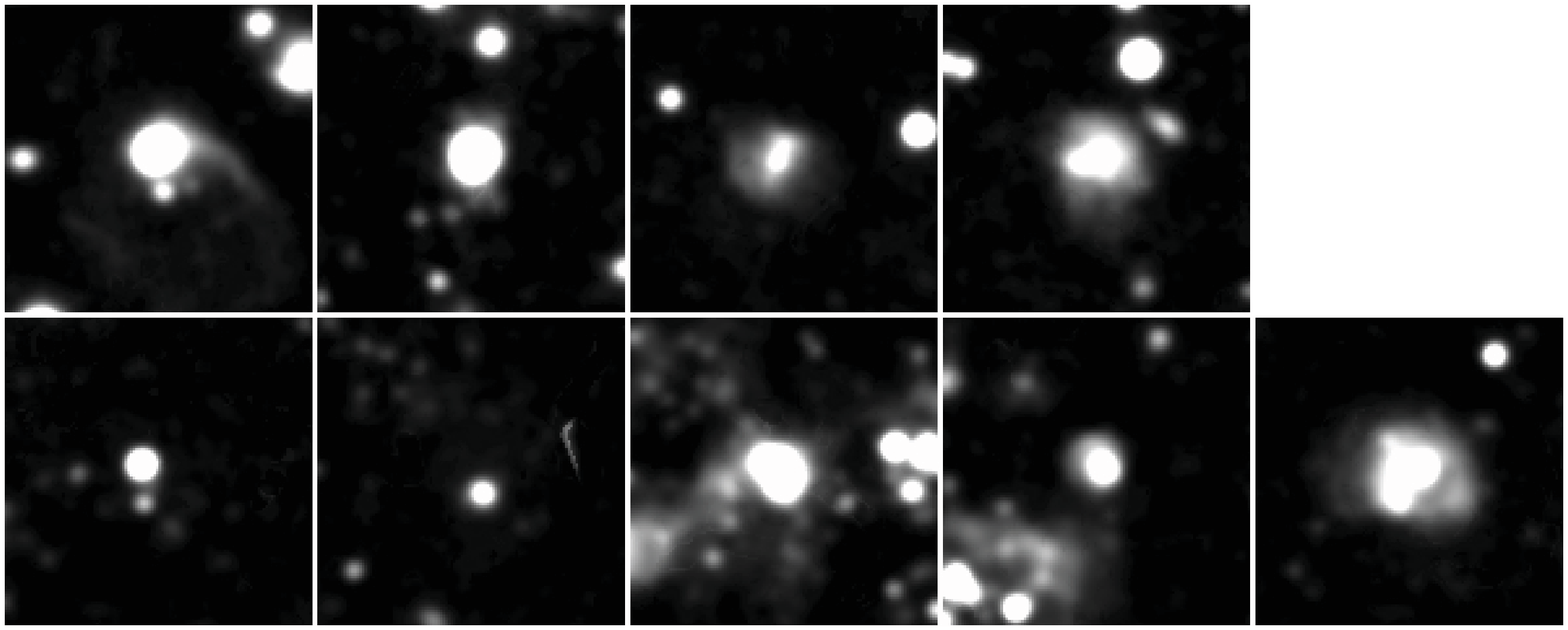}
\caption{$\NA$ images of the $\Halpha$-bright clusters with $\NA-\Rc<-1$ mag and $>10^3\Msun$.
O-type stars must be present to produce this cluster color. The size of each image is $24\arcsec.2\times24\arcsec.2$.
\label{fig:pstamps}}
\end{figure}

The number of blue clusters with FUV-NUV $<$ 0 mag is counted using
Figure \ref{fig:FUVhist} (red) -- after the internal extinction correction,
it is $N_{\rm UV} = 88 \pm 9$.
From Figure \ref{fig:colormag}, $N_{\Halpha} = 9 \pm 3$ for $\NA-\Rc<-1$ mag (see Figure \ref{fig:pstamps}).
The errors are based on the Poisson distribution, and
the counts are made after the internal extinction correction.
Therefore, the observed ratio is $N_{\Halpha}/N_{\rm UV}\sim 0.10 \pm 0.03$.
This is consistent with the prediction of the standard IMF (within a 1$\sigma$ deviation).
The counts of stellar clusters support the standard IMF with a simple aging effect.

It is worth noting that the slight deviation does not likely have a statistical significance.
However, if anything, it is inclined toward the top-heavy IMF side, as opposed to the previous suggestion
of a truncated IMF.

The combination of the parameters we adopt [i.e., low metallicity ($0.2\Zsun$), SMC extinction,
and $A_{\rm V}=0.1$ mag] is the best to consistently explain the new data, model, and previous spectroscopic studies.
The result for the standard IMF does not change even if a higher metallicity ($\sim 0.4\Zsun$), and hence
the LMC extinction curve, is adopted. 
If we ignore the spectroscopic studies, the comparison of the data and model (e.g., Figure \ref{fig:colorcolor})
provides $A_{\rm V}\sim 0.2$ mag. With this extinction, we count $N_{\rm UV} = 86 \pm 9$ and $N_{\Halpha} = 11 \pm 3$
and obtain $N_{\Halpha}/N_{\rm UV}\sim 0.13 \pm 0.04$.
The model ratio is 0.10 with the metallicity $0.4\Zsun$, where  $t_{\rm \Ha} = 5.2$ Myr and $t_{\rm UV} = 51.3$ Myr.
Therefore, the observations and model are consistent even if a different metallicity and extinction are assumed.

\subsection{$\Ha$-to-FUV Flux Ratio}

The $\Ha$-to-FUV flux ratio ($F_{\Ha}/f_{\nu \,\rm FUV}$) over the entire XUV disk is another measure
of the high-mass end of the IMF. We find that the observed ratio is consistent with the one
predicted from the standard IMF.
A lower ratio may suggest a truncated IMF \citep{meu09, lee09},
though an extinction correction introduces an ambiguity \citep{bos09}.
We can estimate the ratio more accurately since we derived the extinction
of $A_{\rm V}=0.1$ mag statistically (\S \ref{sec:intext2}).

In a {\sc Starburst99} simulation with a constant star formation rate, a metallicity of $0.2\Zsun$,
and the Salpeter IMF with ($m_{\rm l}$, $m_{\rm u}$) = (0.1, 100)$\Msun$,
the ratio approaches an approximate constant and is, e.g., $\log (F_{\Ha}/f_{\nu \,\rm FUV}) \sim 13.10$ in unit of $\log({\rm Hz})$
after 1 Gyr assuming a star formation rate of $\rm 1 \Msun/yr$ (though the constant does not depend on these values).
\citet{fum11} calculate the ratio with their own model using the Kroupa IMF and derive $\sim 13.22$ (from their Figure 1).
The scatter between the models is about 30\%.
Compared to these values, the observed ratios previously reported in LSB galaxies by \citet{meu09} -- as low as
one third of the values calculated in these models -- deviate at a significant level with respect to the model uncertainties.

We calculate the $\Ha$ to FUV flux ratio over the XUV disk of M83 by summing up the fluxes of
all  $>3\sigma$ detections in the FUV image (\S \ref{sec:fuvsel}).
Assuming that OUT objects represent the background population (\S \ref{sec:background})
and subtracting their total fluxes form those of IN objects (accounting for the IN/OUT area ratio),
we obtain $F_{\Ha}=8.36\times 10^{28} \,\rm erg\, s^{-1}$ and $f_{\nu \,\rm FUV}=7.23\times 10^{25}  \,\rm erg\, s^{-1} \, Hz^{-1}$.
Adopting $A_{\rm V}=0.1$ mag and the SMC extinction curve (\S \ref{sec:extcor}),
the intrinsic ratio after the extinction correction is $\log (F_{\Ha}/f_{\nu \,\rm FUV})_{\rm int} = 12.93$.
This value is only $\sim 30\%$ lower than our model prediction, and not as much as what the previous study found \citep{meu09}.
As the difference between this value and our model calculation is similar to the scatter between the models,
we consider a truncated IMF to be unwarranted in the case of M83.

\section{Summary}

Using new deep Subaru $\Halpha$ data and  archival GALEX data
of the extended ultraviolet (XUV) disk of M83, we show that
the standard Salpeter IMF explains the number counts of stellar clusters
in the low density, low metallicity environment. Even in clusters with small mass
($10^{2-3}\Msun$) O stars are populated, at least occasionally, suggesting
that the stochastic IMF is preferable to the truncated IMF.
The $\Ha$ to FUV flux ratio over the XUV disk also supports the standard IMF.

The analyses were guided by the stellar population synthesis model {\sc Starburst99}.
The new data, together with the population synthesis model and previous spectroscopic studies,
provide overall
very consistent results at the metallicity ($\sim 0.2\Zsun$), internal extinction
($A_{\rm V}\sim 0.1$ mag), and colors of the stellar clusters in the XUV disk.
We counted the numbers  of FUV-bright clusters (which should have O and/or
B stars) and $\Ha$-bright clusters (having O stars), whose ratio constrains the
high-mass end of the IMF. The ratio of number counts can be translated to the ratio of
their lifetimes with {\sc Starburst99}, under the two assumptions of
instantaneous cluster formation and a constant cluster formation rate in the XUV disk.
As a result, the number counts are consistent with the Standard IMF and a simple aging effect.

This study benefits from the combination of the unparalleled wide field-of-view
and sensitivity of Suprime-Cam.
The new image covers an area far beyond the detection limit of the HI gas (thus,
beyond the boundary of the XUV disk), enabling us to remove the background
contaminations statistically, which had been a difficulty in previous studies.
We found that virtually all blue clusters (FUV-NUV$\lesssim$ 0.0 mag
after the extinction correction; or $\lesssim$ 0.16 mag before the correction) are within the HI gas.

\acknowledgments
We thank Sadanori Okamura and Tomoki Hayashino for providing the redshifted $\Ha$ filter
for Suprime-Cam. JK thanks Peter Capak for his help with the IMCAT software and Deane Peterson and Fred Walter for discussion on stellar spectra.
AGdP is funded by the "Ramo\'n y Cajal" program of the Spanish MICINN and partly supported by the AYA2009-10368 and Consolider-Ingenio 2010 CSD2006-70 projects.
MI is supported by Grants-in-Aid for Scientific Research (no. 22012006, 23540273).
Data analysis were in part carried out on the common use data analysis computer system at the Astronomical Data Center (ADC) of the National Astronomical Observatory of Japan.
Some of the data presented in this paper were obtained from the Multimission Archive at the Space Telescope Science Institute (MAST). STScI is operated by the Association of Universities for Research in Astronomy, Inc., under NASA contract NAS5-26555. Support for MAST for non-HST data is provided by the NASA Office of Space Science via grant NNX09AF08G and by other grants and contracts.

{\it Facilities:} \facility{Subaru}, \facility{GALEX}.

\appendix
\section{Minimum stellar cluster mass to avoid stochasticity}\label{sec:mcl}

We consider the threshold mass of a stellar cluster where the incomplete sampling (filling) of the IMF
(i.e., stochastic effect) becomes important.
It depends on the definition of ``stochastic sampling'', because in reality
an O star with the highest mass $\sim100\Msun$ is rarely found (at least in the Milky Way),
and even a massive cluster (e.g., $10^6\Msun$) may not host any star with $100\Msun$
in a statistical sense.
On the contrary, we could define the fully-sampled IMF as the one in which at least
one O star of any subclass ($>20\Msun$) is populated.
Of course, a cluster is more likely to host a $20\Msun$ star than a $100\Msun$ star,
and the threshold mass should be lower in the former case.
One could also take a different approach;
for example,  \citet{cal10} found $\sim 10^3\Msun$ as the threshold mass where
the $\Ha$ luminosity over cluster mass starts to deviate from a constant.

Here, we take a simpler approach to illustrate a range of threshold cluster mass that
may provide an important guideline. 
The threshold mass necessary to populate the IMF to some upper mass cutoff
depends on that maximum stellar mass

The stellar initial mass function ($\Psi(m)$) with the upper and lower stellar mass cuts ($m_u$
and $m_l$, respectively) is normalized as
\begin{equation}
M_{\rm cl} = \int^{m_u}_{m_l} m \Psi(m) dm,
\end{equation}
where $M_{\rm cl}$ is the total mass of a stellar cluster.
Assuming a power-law IMF ($\Psi(m) = A m^{-\alpha}$) the normalization coefficient ($A$) is 
\begin{equation}
A = M_{\rm cl} \frac{2-\alpha}{m_u^{2-\alpha}-m_l^{2-\alpha}}.
\end{equation}
If we sum up the masses of all stars with masses greater than a threshold stellar mass $m_t$,
the sum is
\begin{equation}
M(>m_t) = \int^{m_u}_{m_t} m \Psi(m) dm.
\end{equation}
In order to populate at least one star as massive as $m_t$, this sum should be greater than $m_t$ 
[i.e., $M(>m_t)>m_t$]; and therefore,
\begin{equation}
M_{\rm cl} > m_t \frac{ m_u^{2-\alpha}-m_l^{2-\alpha} }{ m_u^{2-\alpha}-m_t^{2-\alpha} }
\end{equation}
in the case of a power-law IMF. 
This condition is necessary to {\it unstochastically} populate the IMF up to $m_t$ in the cluster,
providing a threshold cluster mass to have a star up to mass $m_t$.

A similar calculation also provides a guideline of the maximum stellar mass that a cluster with $M_{\rm cl}$
is likely to host -- the IMF would be filled up to $m_t$ if it is populated smoothly from
the lowest stellar mass $m_l$ toward the highest until the total mass reaches the cluster mass $M_{\rm cl}$.
Thus, the maximum stellar mass in the cluster is around $m_t$.

\begin{figure}
\epsscale{0.6}
\plotone{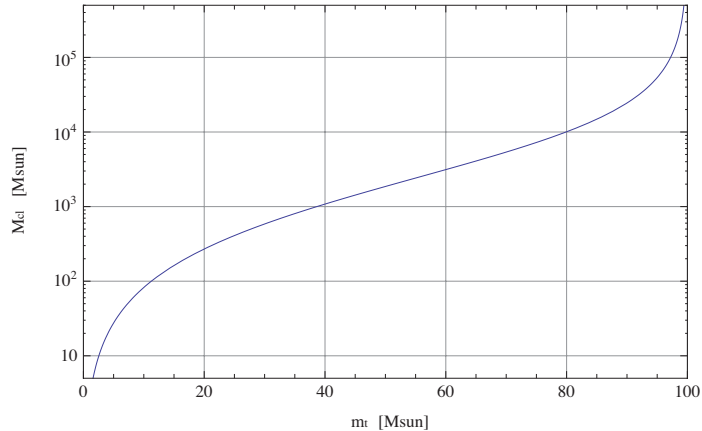}
\caption{Minimum cluster mass to avoid the stochastic sampling effect of the IMF, as a function
of the maximum stellar mass that one wants to populate {\it unstochastically}.
The cluster mass needs to be $10^3$, $10^4$, and a few $10^5\Msun$ to populate
an O star up to 40, 80, 100$\Msun$, respectively.
\label{fig:mcl}}
\end{figure}

Figure \ref{fig:mcl} shows the threshold cluster mass $M_{\rm cl}$ as a function of the
maximum stellar mass $m_t$ that should be populated.
We assumed the Salpeter IMF ($\alpha=2.35$), $m_l=0.1\Msun$, and $m_u=100\Msun$.
In order to have a $40\Msun$ star the cluster mass should be at minimum $10^3\Msun$,
while to have $80\Msun$ it needs to be $10^4\Msun$. All stars with $>20\Msun$ are
classified as O-type stars; and to have at least one O star the cluster mass should be
greater than $\sim 400\Msun$.
To fill the IMF literally up to $\sim 100\Msun$, the cluster mass
should be on the order of $10^5\Msun$. If we adopt $50\Msun$ (arbitrarily) as a typical mass of O stars,
the cluster mass should be $>2000\Msun$ to fully populate the IMF up to this typical O star mass
and avoid the stochastic effect.

These threshold cluster masses are meaningful only in a statistical sense.
A stochastic effect could appear even above this threshold mass
if a Monte-Carlo simulation is performed.
Even if a cluster mass is above the threshold and the cluster {\it can} host a star with $m_t$,
this mass could be split and allocated to multiple stars with smaller masses
(whose total is $m_t$).
On the other hand, a cluster with mass below the threshold could occasionally have a star
with mass $>m_t$ at the expense of many less massive stars.
The threshold cluster mass should be taken as a minimum mass to avoid the stochastic effect,
only in a statistical sense.
Despite this limitation, our estimate provides a crude guideline
for discussion of the stochastic sampling of the IMF.

\section{UV Spectra of O, B and A stars}\label{sec:starspec}

\begin{figure}
\epsscale{0.9}
\plotone{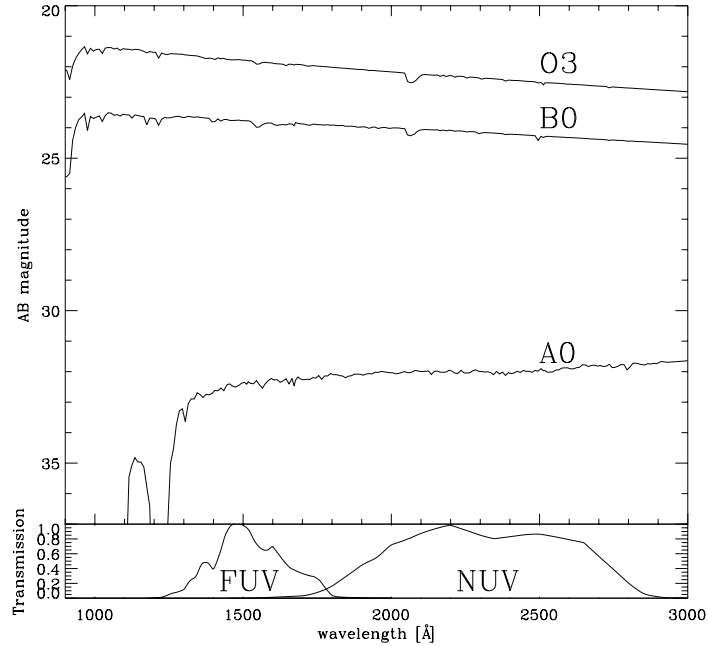}
\caption{{\it Top:} UV spectra of O3 ($87.6\Msun$), B0 ($17.5\Msun$) and A0 ($2.9\Msun$) stars.
The $y$-axis is an apparent magnitude at the distance of M83 ($d=4.5{\rm \, Mpc}$; $m-M$=28.27 mag).
{\it Bottom: } transmission curves of GALEX FUV and NUV bands.
\label{fig:starspec}}
\end{figure}

The spectra of O3, B0, and A0-type stars from the Kurucz's Atlas9 grid \citep{kur92} are plotted in Figure \ref{fig:starspec}.
The masses of these stars are approximately 87.6, 17.5, and 2.9$\Msun$, respectively \citep{ste03, cox99}.
We adopt a metallicity of $\rm [Fe/H]=-0.5$ (or $\sim0.3\Zsun$) which is available in the grid.
The effective temperature ($T_{\rm eff}$) and gravity ($\log g$) are required
to find a model from the grid, and the stellar radius $R$ is necessary to calculate the total flux.
From the Atlas9 grid, we choose the models whose parameter sets are the closest to
the stellar parameters presented in \citet{ste03} or \citet{cox99}:
($T_{\rm eff}$, $\log g$, $R$) =
(50000 K, 5.0, 13.2 $\Rsun$), (30000, 4.0, 7.4), and (10000, 4.0, 2.4) for O3, B0, and A0, respectively.
We note that the spectra of high-mass stars are not calibrated very well as only few of those exist in
the solar neighborhood for calibration.
In addition, an adjustment of the parameters is necessary to fit observed spectra of individual stars \citep{ste03, lei10}.
The spectra that we present here are only a reference to understand our observations.

The continuum slopes are more or less the same between the O and B stars in the wavelength range of GALEX bands.
On the other hand, the A star spectrum is rising from FUV to NUV.
The FUV absolute magnitude (AB) and FUV-NUV color of the individual stars for three metallicity
[Fe/H]=+0.0 ($1.0\Zsun$), -0.5 ($0.3\Zsun$), -1.0 ($0.1\Zsun$) are in Table \ref{tab:uvstar}.
We also list the Lyman continuum photon emission rate ($Q_{\rm Lyc}$),
which can be used to calculate an H$\alpha$ luminosity
($L_{\Ha} {\rm [erg \, s^{-1}]}= 1.37 \times 10^{-12}Q_{\rm Lyc} {\rm [s^{-1}]}$ in case B recombination).
The $Q_{\rm Lyc}$ is calculated by integrating
the model spectra over the frequency above the ionizing frequency (i.e., the wavelength of $912\AA$).
These $Q_{\rm Lyc}$ values are 4-5 times smaller than those in \citet{ste03}, reflecting the uncertainty
in model spectra of high-mass stars.

\begin{deluxetable}{cccccccccc}
\tablecolumns{10}
\tablewidth{0pc}
\tablecaption{UV Color of High-mass Star Models}
\tablehead{
\colhead{}    & \colhead{}    & \multicolumn{2}{c}{$\rm [Fe/H]=+0.0$}   & \colhead{} & \multicolumn{2}{c}{-0.5}  & \colhead{} &  \multicolumn{2}{c}{-1.0}  \\
 \cline{3-4} \cline{6-7} \cline{9-10}\\
\colhead{Type} & \colhead{$\log Q_{\rm Lyc}$}    &\colhead{FUV}  & \colhead{FUV-NUV} & \colhead{} & \colhead{FUV}  & \colhead{FUV-NUV} & \colhead{} & \colhead{FUV}  & \colhead{FUV-NUV} \\
\colhead{} & \colhead{($\rm s^{-1}$)}   & \colhead{(mag)}  & \colhead{(mag)} & \colhead{} & \colhead{(mag)}  & \colhead{(mag)} & \colhead{} & \colhead{(mag)}  & \colhead{(mag)}}
\startdata
O3 & 49.16 & -6.47& -0.54 && -6.46 & -0.57 &&  -6.44 & -0.58 \\
B0 &  47.37 & -4.44 & -0.31 &&  -4.47 & -0.37 && -4.48 & -0.41 \\
A0 &  37.24 & 4.30 & 0.58 && 4.16 & 0.45 &&  4.10 & 0.40
\enddata
\tablecomments{FUV magnitudes are the absolute magnitude.}
\label{tab:uvstar}
\end{deluxetable}


\end{document}